\documentclass{osa-article}

\journal{josab}

\articletype{Research Article}

\usepackage{times}

\newcommand{\redline}[1]{{#1}}

\begin{document}

\title{Toward A Reconfigurable Quantum Network Enabled by a Broadband Entangled Source}

\author{Eric Y. Zhu,\authormark{1,*} Costantino Corbari,\authormark{2} Alexey Gladyshev,\authormark{3} Peter G Kazansky,\authormark{2} Hoi-Kwong Lo,\authormark{1,4} and Li Qian\authormark{1,5}}

\address{
\authormark{1}Dept. of Electrical and Computer Engineering, University of Toronto, 10 King's College Rd., Toronto, ON M5S 3G4, Canada\\
\authormark{2}Optoelectronics Research Centre, University of Southampton, SO17 1BJ, United Kingdom\\
\authormark{3}Fiber Optics Research Center of the Russian Academy of Sciences, 119333, 38 Vavilov street, Moscow, Russia\\
\authormark{4}Dept. of Physics, University of Toronto, 60 St. George St., Toronto, ON M5S 1A7, Canada\\
}

\email{\authormark{*}eric.zhu@utoronto.ca} 
\email{\authormark{5}l.qian@utoronto.ca} 



\begin{abstract}
We present a \redline{proof-of-principle} experimental demonstration of a reconfigurable entanglement distribution scheme utilizing a poled fiber-based source of broadband polarization-entangled photon pairs and dense wavelength-division multiplexing (DWDM). The large bandwidth (> 90 nm centered about 1555 nm) and highly spectrally-correlated nature of the entangled source can be exploited to allow for the generation of more than 25 frequency-conjugate entangled pairs when aligned to the standard 200-GHz ITU grid. In this work, 3 frequency-conjugate entangled pairs are used to demonstrate quantum key distribution (QKD), with wavelength-selective switching done manually. The entangled pairs are delivered over 40 km of actual fiber, and an estimated secure key rate of up to 20 bits/s per bi-party is obtained. 
\end{abstract}

\section{Introduction}

Quantum key distribution (QKD) offers, in principle, an unconditionally secure method for two parties to generate a private cryptographic key.  Many point-to-point (PTP) demonstrations using weak-coherent source-based \cite{DecoyQKD144km} and entanglement-based \cite{ursin2007entanglement144km,treiber2009fully} quantum-cryptographic protocols such as the vaunted BB84 \cite{BB84paper},  BBM92 \cite{BBM92}
and E91 \cite{EkertEntanglementQKD} (respectively) have been demonstrated.  
However, PTP connections do not provide an efficient method to connect multiple users, 
and efforts thusfar in extending PTP links into multi-user networks have proven to be 
cumbersome at best.

To that end, QKD networks based upon the hub-and-spoke model, where many end-users can be connected to a trusted node \cite{elliott2002building,SECOQCVienna2009,TokyoQKD2011,hughes2013network}, are used.  Should any one end-user wish to communicate with another user secretly, a random secret key  is first generated at the node, 
then QKD is 
performed between the node and each user separately; 
finally, the key is sent to each end-user classically using the QKD key material as 
a one-time pad. 
However, the trust model for such a network topology is inherently problematic, as any security vulnerability at the node will compromise the entire network.  
Additionally, time-multiplexing is often required in such circumstances to service multiple users \cite{Choi1300QKD,frohlich2013quantum}, and only one end-user can perform QKD with the central node at a time.  

Other schemes \cite{ChenStarNetworkQKD,USTC5nodeQKD} involving wavelength-division multiplexing (WDM) have been introduced where individual users on a PTP network can perform QKD  with one another  simply by addressing each other at different laser wavelengths.  
All users are equivalent on the network, and no central trusted node is required.  
However, each user must have both single photon detectors and laser sources
at their disposal, and the incremental cost of adding a new user to an $N$-user network 
requires increasing the number of laser wavelengths available to each user to $N+1$.  

\begin{figure}[t!]
	\centering
	\includegraphics[width=8cm]{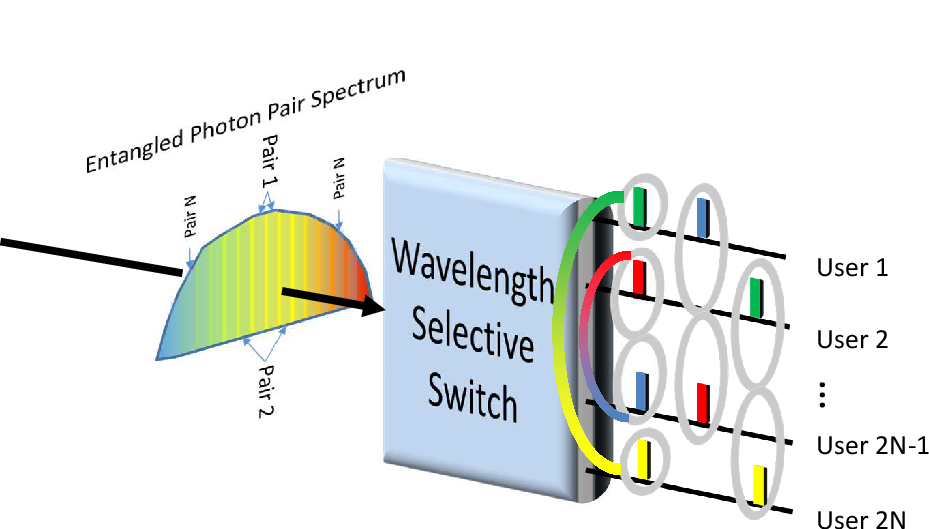}
	\caption{\label{fig:StarNet}
	The use of a broadband entangled source and a wavelength-selective switch 
	(WSS) allows for any two parties on the network to receive
	frequency-conjugate polarization-entangled photon pairs.
	The different figure-eights denote the use of different frequency-conjugate 
	pairs to distribute entanglement to different users.  
	The broadband nature of the source also allows for multiple simultaneous connections.  
	}
\end{figure}

In contrast, we show that a reconfigurable quantum network utilizing 
the distribution of polarization-entangled photon pairs from a (potentially untrustworthy) central office can be realized efficiently using a single 
broadband entangled source (Fig. \ref{fig:StarNet}).  
While previous works \cite{OrieuxQST:2016, LimPEPPdistribute10km,Herbauts:13} have
 demonstrated  the distribution of polarization-entangled photon pairs at distances 
 up to 50 km of fiber using WDM technology, the entangled sources used in those cases left much to be desired.  
 The generation of entangled photon pairs relied upon interferometric methods in \cite{LimPEPPdistribute10km,Herbauts:13} which added to experimental 
 complexity, or were limited in both bandwidth \cite{Herbauts:13,OrieuxQST:2016} (to around 10 nm) 
 and quality of entanglement (raw entanglement fidelity $F<$0.94).
Meanwhile, the fiber-based source \cite{zhueyiOL2013,AlexChenOE2017} used in this work has a bandwidth of over 90 nm, with a quality of 
entanglement that remains undiminshed ($F>$0.98) over the entire band.  

Due to this large bandwidth, the polarization-entangled photon pairs from our source are also highly spectrally-anticorrelated, which allows for multiple ($N$) frequency-conjugate pairs to be carved out of the spectrum and distributed to $N$ pairs of users simultaneously. 
When 200-GHz-spacing dense WDM (DWDM) filters are used, $N=25$ is achievable with our source.  
%
With a $ 2N\times 2N$ wavelength-selective switch (WSS) to perform dynamic spectrum allocation \cite{oh2011coincidence}, 
and without additional hardware resources, any user on the 
 network can perform QKD with any other user.
Additionally, our scheme allows  multiple users ($2N$)  to  perform 
QKD simultaneously; when combined 
with time-multiplexing, such a scheme can extend the number of (non-simultaneous) users well beyond $2N$.  
  
\redline{As a proof of principle, we choose 3 frequency-conjugate pairs out of the 25 to demonstrate this entanglement distribution idea.}  With two single-photon detectors at our disposal (one detector for Alice, the other for Bob), filter pairs are swapped manually to demonstrate dynamic wavelength allocation.

%
\section{Experimental scheme}
\redline{
In this section, we first provide an overview of the entangled source, the poled optical fiber, followed  by a description of the experimental setup used for entanglement distribution and QKD.  
\newline
}

\begin{figure}[h!]
	\centering
	\includegraphics[width=8cm]{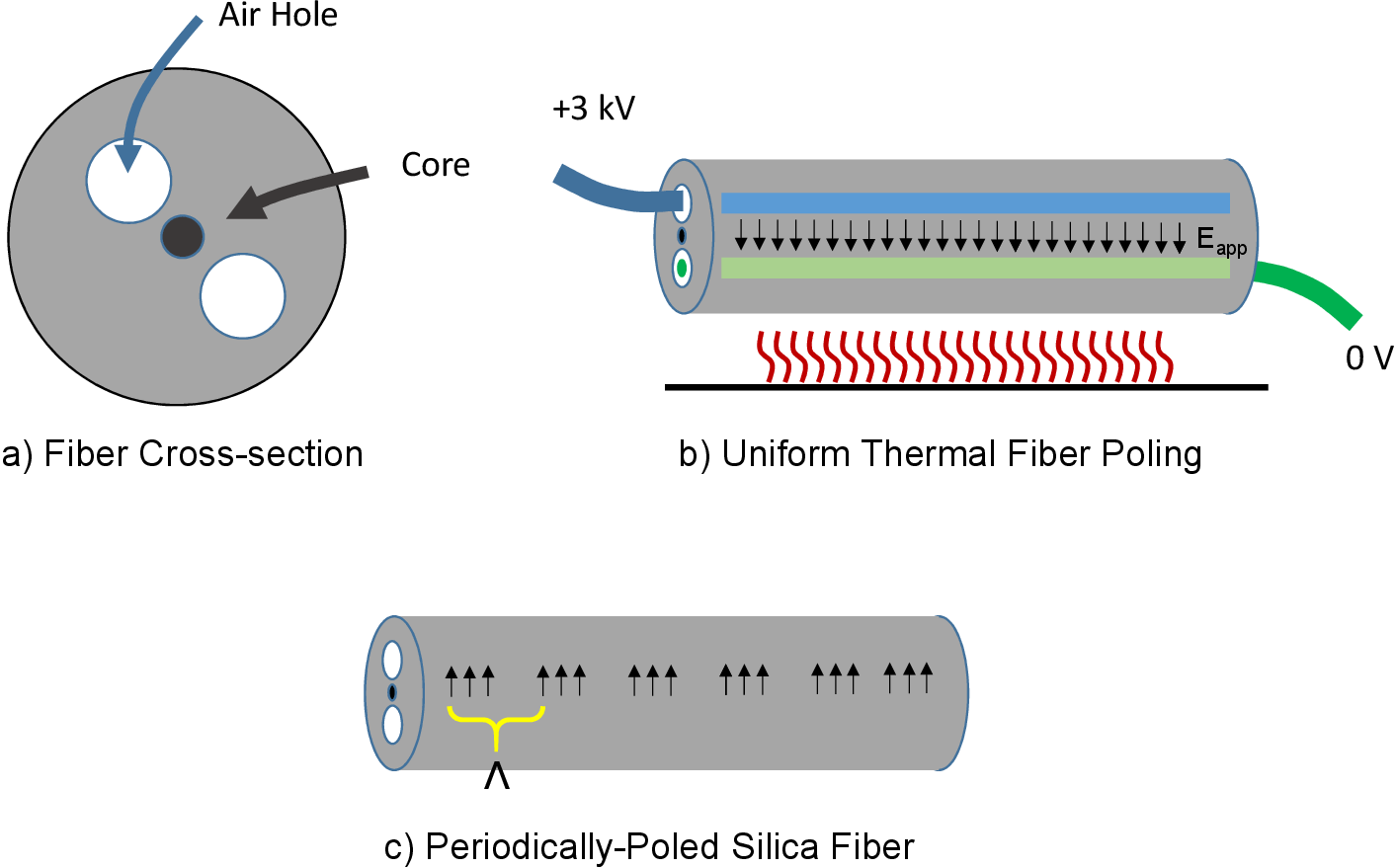}
	\caption{ \label{fig:PPSF}
\redline{	
	(Color online)
	a)  Cross-section of a twin-hole fiber used for thermal poling.  Two large air holes sandwich the fiber core.
	b) The thermal poling process.  Electrodes are threaded through the air holes, and a large DC potential ($E_{app}$) is applied across them while the fiber is heated.  After the poling process, the electrodes are removed,  and a frozen in electric field $E_{DC}$ (whose polarity is opposite to the poling field $E_{app}$) is created along the length of the fiber. 
	c) Quasi phase-matching is achieved by periodically erasing the $E_{DC}$ (and consequently, the SON) along the length of the fiber with UV light.  The fiber is now a periodically-poled silica fiber (PPSF).  
		} 
	}
\end{figure}

\redline{
The source of polarization-entangled photon pairs (PEPPs) used in this work is based upon the poled optical fiber 
\cite{kazansky1994thermally,Canagasabey:09high}.  It is a step-index fiber composed of fused silica glass that exhibits a non-zero second-order nonlinearity (SON).  The SON is induced through the process of thermal poling
 \cite{kazansky1994thermally}, and involves the application of a large DC electric field (kV/$\mu$m) at elevated  temperatures (500 K) that breaks the inversion symmetry of the glass by creating a frozen-in electric field   \cite{kazansky1994thermally,Canagasabey:09high} $E_{DC}$; this field in turn induces a second-order
  nonlinearity (SON): $\chi_{eff}^{(2)}=3\chi^{(3)} E_{DC}$ \cite{zhueyi2010OL}.  The poling process is facilitated by the geometry of twin-holed fiber \cite{Canagasabey:09high}, where two 
    air-holes [Fig. \ref{fig:PPSF}(a)] run along the length of the fiber and sandwich the fiber core; these holes allow for thin 
     electrodes to be threaded through the length of the fiber to allow for the application of the large DC electric 
      field [Fig. \ref{fig:PPSF}(b)].  The non-zero SON in the fiber then enables the generation of photon pairs through the process of spontaneous parametric downconversion (SPDC).  
}

\redline{
However, while a SON is a necessary ingredient for efficient SPDC, it is not sufficient.  Phase-matching between pump and downconverted photon pairs must also be satisfied.  This is done by periodically erasing \cite{corbari2005aff} the SON [Fig. \ref{fig:PPSF}(c)] along the length of the fiber (at a period of $\Lambda$) with ultraviolet (UV) light, which yields a quasi phase-matching (QPM) condition:
\[
k(\omega_p )= k(\omega_s)+k(\omega_i)+\frac{2\pi}{\Lambda}, 
\]
where $k(\omega_m)$ is the fiber propagation constant at frequency $\omega_m$, with $m$ being either the pump ($p$), signal ($s$), or idler ($i$).  The value of $\Lambda$ can be judiciously chosen so that the pump, signal, and idler are each at an arbitrary wavelength within the transparency window of the fiber, subject to the energy conservation constraint $\omega_p= \omega_s+\omega_i$.  In our case, we choose $\Lambda$ so that the downconverted signal and idler photons are generated in the 1.5-micron telecom band.  
After periodic UV erasure, the uniformally-poled twin-holed optical fiber has now become a periodically-poled silica fiber (PPSF).  We will use PPSF and `poled fiber' interchangeably in the remainder of the text.  
}

\redline{
The presence of a type-II phasematching \cite{zhueyi2010OL} and the extremely low birefringence of the fiber (< 10 fs/m) allow for the direct generation of PEPPs \cite{zhueyiPRL2012} in the triplet state
\[
|\Psi^{+}\rangle = \frac{1}{\sqrt{2}}\left(|HV\rangle +|VH\rangle\right).
\]
 Additionally, due to the low group velocity dispersion of the poled fiber, PEPPs can be generated over an
  extremely broad band of more than 90 nm \cite{zhueyiOL2013,AlexChenOE2017} without significant reduction in the quality    of polarization entanglement.    
}

\begin{figure}[h!t]
	\centering
	\includegraphics[width=8cm]{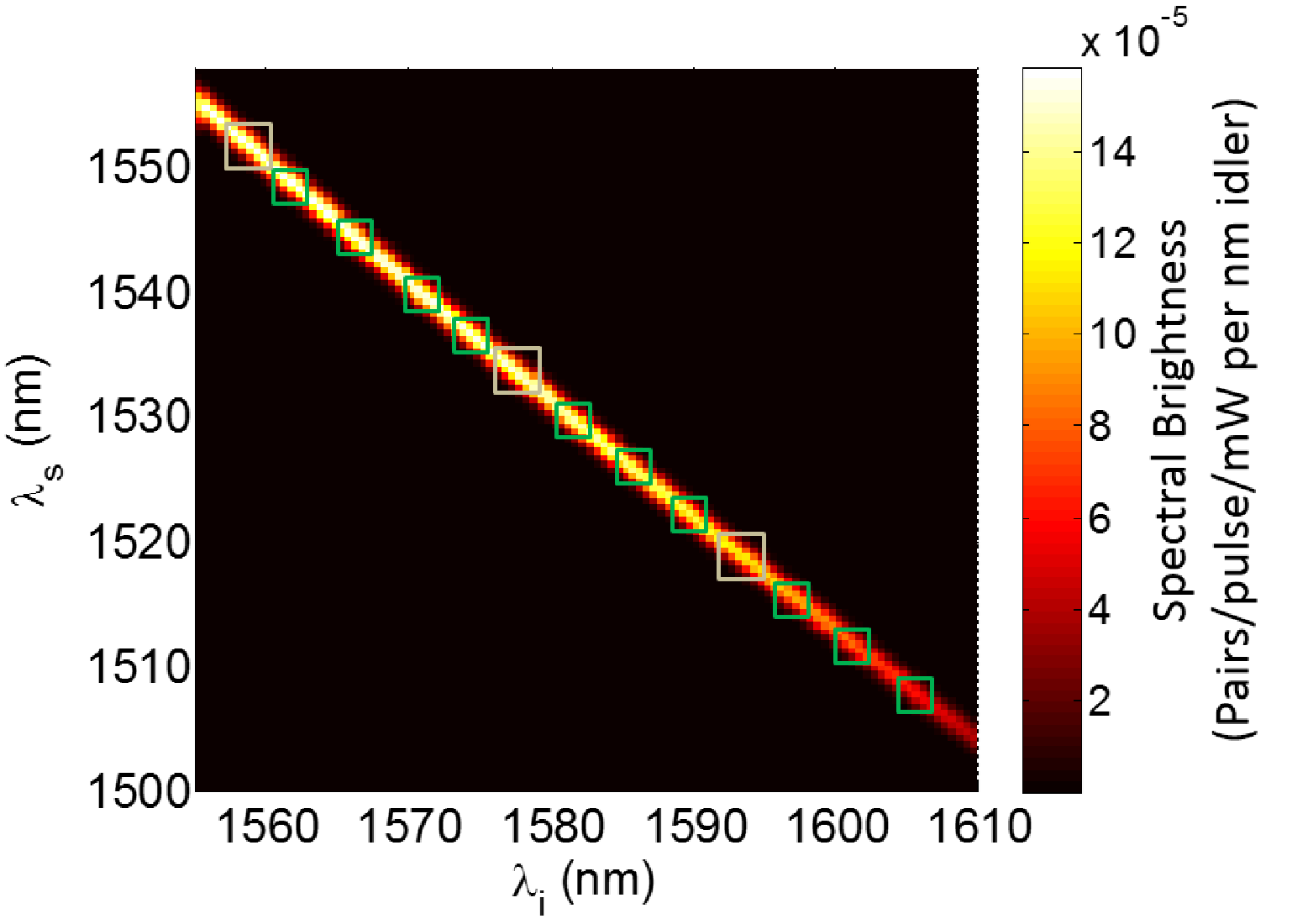}
	\caption{\label{fig:PPSF2014_JSA} 
	(Color online)		
	The joint spectral intensity (JSI) of the downconverted photons is plotted as a function of 
	the signal and idler wavelength.  
	A total of $6.8\times10^{-3}$ pairs/pulse per mW of average pump power is generated over 
	the entire spectrum.  
	We observe that the JSI is extremely spectrally-anticorrelated.  
	This anti-correlation allows us to separate the spectrum into many frequency-conjugate pairs (green and white square boxes).  
	The three pairs used in this work are highlighted by larger white boxes.  See Table \ref{tab:EntResults} for the 
	central wavelengths of each pair.  
	}
\end{figure}

A modified regeneratively-modelocked Ti:Sapphire laser (Spectra-Physics Tsunami) set to 777.45 nm  and stabilized with an intracavity interference filter is used to pump the poled fiber with a 81.6 MHz train of 400-ps-long pulses; the average power inside the poled fiber is approximately 50 mW.  
The broadband spectral correlations \cite{zhueyiOL2013} (Fig. \ref{fig:PPSF2014_JSA}) 
of the source are employed to 
distribute multiple frequency-conjugate pairs centered about 1554.9 nm (= $2\times 777.45$ nm) to different parties.   
This central wavelength also happens to coincide with Channel 28 of the ITU DWDM grid \cite{web:ITUdwdm}.

Figure \ref{fig:PPSF2014_JSA} shows the joint-spectral intensity (JSI) of the 
downconverted photon pairs from the poled fiber.  The measurement of the JSI was obtained with a fiber 
spectrometer \cite{SilberHorn2009,zhueyiFiberSpec}.  
Due to the highly 
spectrally-anticorrelated nature of the downconverted photons, 
many frequency-conjugate pairs can be obtained simultaneously using
dense wavelength division multiplexing (DWDM) technology.

\begin{table}[ht!]
{
		\caption{
Polarization Entanglement Figures of Merit		
		\label{tab:EntResults}
		} 	
	\begin{center}
	\begin{tabular}{  c   c   c    c }
	\hline
		\centering	
		$\lambda_s$ (nm) 	& 	1553.3 	&		1533.3 			&		1518.7 \\
		$\lambda_i$ (nm)		&		1556.6 	&		1577.1 			&		1593.0 \\ \hline 
	 Tangle 	   & 		0.967$\pm$0.002   									& 0.964$\pm$0.004  								& 0.952$\pm$0.008\\
	 			       \hline

	 Fidelity to $\Psi^+$   & 	0.989$\pm0.001$					& 	0.989$\pm{0.001}$			& 0.985$\pm{0.002}$\\
\hline 
\end{tabular}
\end{center}
}
\end{table}

In Table \ref{tab:EntResults}, the figures of merit (tangle, fidelity) 
of the polarization entanglement 
for three frequency-conjugate pairs (with signal-idler separation varying from 3 nm, 45 nm, and 75 nm) are given.
The three wavelength pairs were chosen not only to demonstrate the large bandwidth of the source, but also
to show that the quality of entanglement is consistently high over the entire band.  
The values in Table \ref{tab:EntResults} were obtained by performing quantum state tomography at a pair generation rates of approximately $4\times10^{-3}$ pairs/pulse/nm using DWDM filters with 1 nm bandwidths.  
Each of these pairs is highlighted in Fig. \ref{fig:PPSF2014_JSA} with a white box.   
While only 3 distinct conjugate pairs are used here, more than 
25 bi-parties can be accomodated  inside this downconversion bandwidth 
of 
the poled-fiber, assuming channel spacings  of 200 GHz that are locked to the ITU grid \cite{web:ITUdwdm}.  
An even greater number of biparties can be accommodated when the channel 
spacings are further reduced to 100 or 50 GHz. 
\newline

\begin{figure*}[t]
\centering
	\includegraphics[width=8cm]{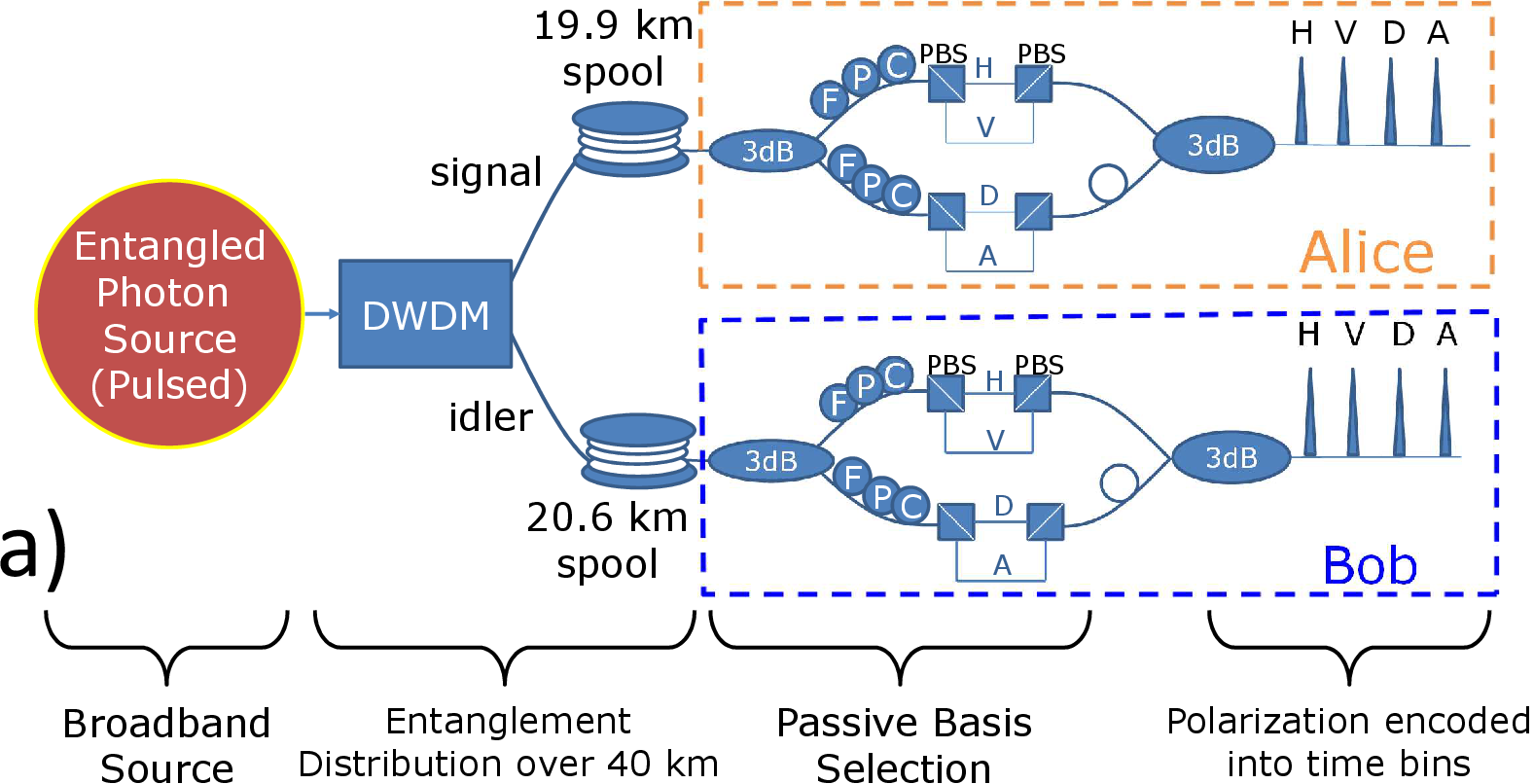}
	\newline
	\newline
	\includegraphics[width=8cm]{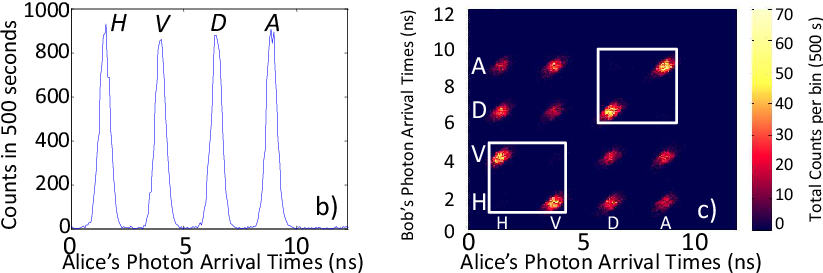}
	\caption{\label{fig:QKDsetup}
	(a)  The full experimental setup for QKD.  The
	signal and idler photons are separated in wavelength by a DWDM filter before they each
	traverse a 20-km fiber spool and arrive at Alice and Bob's polarization analyzers.  
	(b)  	During a QKD experiment, Alice tabulates the number of detector `clicks'
	she receives as a function of their arrival times (with respect to the pump laser sync 
	signal).  
	(c)
	A two-dimensional histogram displaying all the coincidence counts observed
	between Alice and Bob.  In a real QKD implementation, only the counts  
	for which Alice and Bob's bases coincided  would be kept; these regions
	are highlighted by square boxes.  The data shown in (b) and (c) are \redline{obtained} 
	when the signal/idler WDMs are centered about 
	1553.3/1556.65 nm, and a pair generation of $8.9\times10^{-3}$ pairs/pulse.  
	The resolution of the histogram is 64 ps.  	
	\vspace{-5pt}
	}	
\end{figure*}

In our proof-of-principle demonstration of multi-user QKD, 
the experimental setup shown in Fig. \ref{fig:QKDsetup}(a) is used.  
A cascaded set of DWDMs is employed 
to spectrally separate the signal from the idler.  
Each photon then traverses a 20 km 
spool of SMF-28 before reaching the polarization analyzers.  
The BBM92 \cite{BBM92} protocol is implemented, and experiments are run 
with three different frequency-conjugate signal/idler wavelength pairs 
(Table \ref{tab:QKDresults}).  

Switching between the various sets of frequency-conjugate filters is done manually, with
all other conditions (such as the fiber spans, detectors, pump power, and pump wavelength)  kept the same; this allows us to make a meaningful performance comparison between the various wavelength pairs used.  
The use of a wavelength-selective switch would be a trivial addition to the scheme.  

{
 All three filter sets used have very steep cut-on and cut-off wavelengths ($>$ 17 dB/nm), and allow for more than 30 dB extinction between adjacent DWDM channels, meaning crosstalk between adjacent quantum DWDM channels would be negligible.  
The polarization-dependent loss for each DWDM is  also negligible ($<$ 0.1 dB), as is the polarization mode dispersion (PMD $<$ 0.1 ps); this means that the polarization entanglement of the photon pairs does not degrade after passing through these filters.
}

%


\begin{table}
{	                                              
\caption{ QKD Experimental Parameters and Results \label{tab:QKDresults}}  
	\begin{center}	                
			\begin{tabular}{  c   c   c    c }
				\hline
					\centering	
					$\lambda_s$ (nm) 	& 	1553.3 	&		1533.3 			&		1518.7 \\
					$\lambda_i$ (nm)		&		1556.6 	&		1577.1 			&		1593.0 \\ \hline 
	                                                          
		Pairs/pulse				&											&									&					\\
		Generated	per	 &			$8.9\times10^{-3}$				&	$7.6\times10^{-3}$ 				&		5.2$\times10^{-3}$	\\
		 channel &											&									&				\\ \hline
		System Loss &    				&									&						\\ 
		Signal+Idler:   				&    19.4+19.5			&		18.7+19.8	&		18.7+20.7 	\\  
		(dB)  						&						&					& 			\\ \hline

		Sifted Key				&							&									&					\\ 
		Rate (bits/s)				&	32.5$\pm0.7$                  &			29.4$\pm0.7$                  & 		16.3$\pm0.5$ 		\\ 
		($\pm 3\sigma$)                &							&									&					\\ \hline
		QBER ($\pm 3\sigma$):   	& 							& 									& 						\\
		$e_H\ (\%)$				&			$2.35\pm0.51$	&			$1.68\pm0.45$				&		$4.72\pm1.0$		\\ 
		$e_D\ (\%)$				&			$2.15\pm0.48$	&			$2.23\pm0.51$				&		$7.22\pm1.2$			\\ \hline
		Secure Key				&							&									& 						\\
		Rate (bits/s)				&	$20.5\pm1.0$				& 			$19.7\pm1.0$						& 	$4.0\pm0.9$				\\ 
		($\pm 3\sigma$)			&							&									&							\\	\hline
			\end{tabular}		
 	\end{center}
	}
	\vspace{-15pt}
\end{table}

Each polarization analyzer used in the  experiment (Fig. \ref{fig:QKDsetup}a)
consists of a 50/50 (3 dB) beam splitter that 
passively selects the basis in which the photon will be measured.  
\redline{
Passive basis selection is used here to reduce
the complexity of the system and allow for higher speeds, in comparison to active basis selection.  }
In the top arm of the analyzer, the $H/V$ basis is measured with an unbalanced Mach-Zehnder interferometer (MZI) built from two fiber-pigtailed polarizing beam splitters (PBS); this allows for a time-multiplexed measurement scheme in which 
a $V$-polarized photon will arrive 2.5 ns later than an $H$-polarized photon \cite{zhueyiOL2013}.  
A fiber-based polarization controller (FPC) is placed before the PBS-MZI to allow for proper polarization alignment. 
In the lower arm of the analyzer, a similar unbalanced MZI setup is used to measure the photon
in the $D/A$ basis.  
The two arms are then recombined (at a loss penalty of 3 dB) so that all measurement outcomes $H/V/D/A$ can be discriminated in time and measured by a single detector 
(the idQuantique id220 free-running single photon detector [SPD])
for each party.  This results in the mapping of the polarization degree-of-freedom onto timebins that are evenly-spaced by 2.5 ns [Fig. \ref{fig:QKDsetup}(b)].

\redline{
An electronic synchronization ('sync') signal derived from the pump laser is used as the global timing reference.  A commercial time-to-digital converter, the PicoQuant HydraHarp 400, running in time-tagged T3 mode, is used to generate a two-dimensional histogram, which graphically tallies the coincidence detections as a function of 
signal (Alice) and idler (Bob) arrival times [Fig. \ref{fig:QKDsetup}(c)]  with respect to the sync signal.  This is done at a resolution of 64 ps. For each party, measurements are then binned [Fig. \ref{fig:QKDsetup}(b)] into 1-ns-wide time slots corresponding to the 4 measurement outcomes ($H$, $V$, $D$, and $A$). 
}

\section{Results and Discussion}

\redline{
Table \ref{tab:QKDresults} summarizes the results of representative QKD experiments performed for the 3 different
frequency-conjugate pairs.  The acquisition time  $T_\textrm{acq}$  for each set of data is 500 seconds.  
}

The estimated QBER (quantum bit error rate) is calculated from the histogrammed data [Fig. \ref{fig:QKDsetup}(c)].  
As an example, the QBER in the $H/V$ basis is calculated using the formula:
\begin{equation}\label{eq:CalculateQBER}
	e_H = \frac{C_{H,H} + C_{V,V}}{C_{H,H} + C_{V,V} + C_{V,H} + C_{H,V}},
\end{equation}
where $C_{P_A,P_B}$ is the number of coincidence counts observed 
when Alice and Bob measure in the $P_A$ and $P_B$ polarizations (respectively).  
 
The sifted key rate $R_{\textrm{sif}}$ is calculated using the formula:
\begin{equation}\label{eq:CalculateSiftKey}
	R_{\textrm{sif}} = \frac{C_{H,V} + C_{V,H} + C_{D,D} + C_{A,A}}{T_{\textrm{acq}}}.
\end{equation}
We note that a bit-flip must be performed for the $H/V$ coincidences [Fig. \ref{fig:QKDsetup}(c), bottom left corner] to account for the anti-correlated outcomes when the Bell state $|\Psi^{+}\rangle$ is used for QKD.

The secure key rate $R_{\textrm{sec}}$ 
is estimated from the formula \cite{gottesman2002security}:
\begin{equation}
\label{eq:Rsec}
	R_{\textrm{sec}}= R_{\textrm{sif}} \biggl(	  1 - \frac{1}{2}\left[h(e_H^u) + h(e_D^u)\right]   
									  -\frac{1}{2}\left[f(e_H) h(e_H) + f(e_D) h(e_D)\right]	\biggr), 
\end{equation}
%
\noindent
where the factor of $\frac{1}{2}$ is due to the passive basis selection being unbiased, $h(x)$ is the 
binary Shannon entropy ($h(x) = -x \log_2(x) - (1-x)\log_2 (1-x)$), $e_H^u$ ($e_D^u$) is 
the upper bound for the QBER in the $H/V$ ($D/A$) basis  used to estimate the number of keys 
lost due to privacy amplification \cite{LutkenhausPrivacyAmp2000}, 
and $f(e)$ ($\geq 1$, but taken to be 1.2 hereafter) 
is a measure of the efficiency of
the error-correcting codes \cite{brassardErrorCorrection}.

Finite key analysis is summarized in Table \ref{tab:QKDresults}, with the values of QBER and $R_{\textrm{sec}}$ 
accompanied by their uncertainties at 3 standard deviations ($\pm 3\sigma$).  
{If we assume the fair-sampling assumption holds, 
a low QBER is enough to ensure the  security of the key given an untrustworthy central office \cite{gottesman2002security}. }

\begin{figure}[h!]
\centering
	\includegraphics[width=8cm]{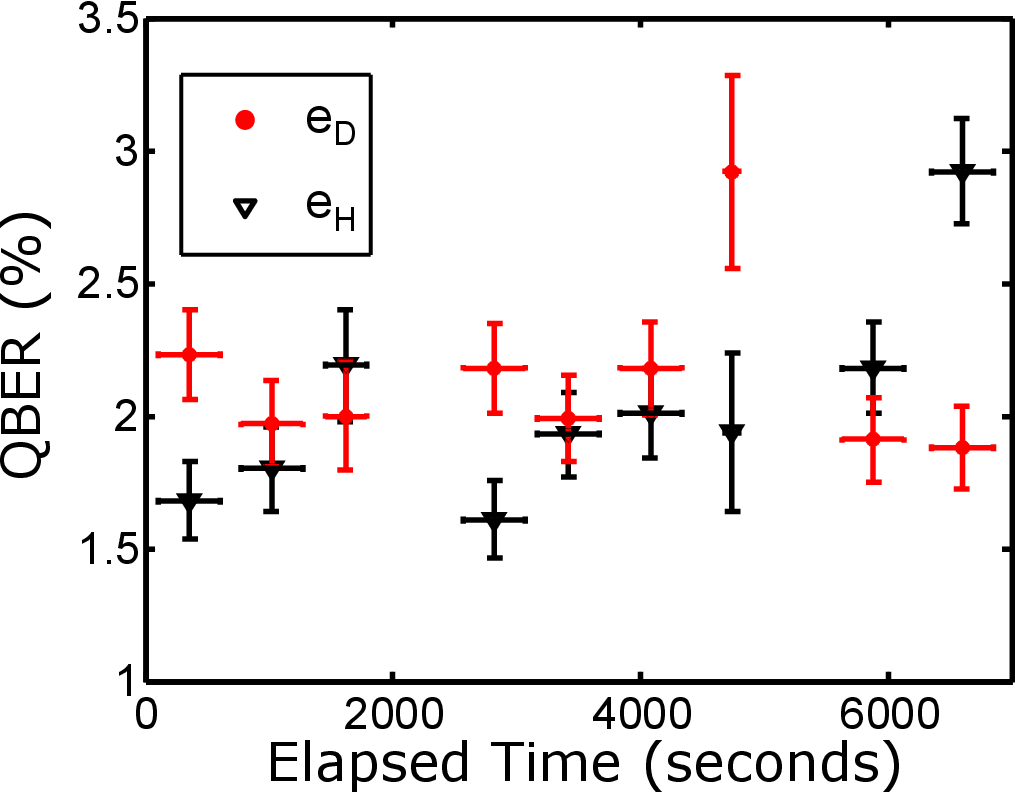}
	\caption{\label{fig:QBErplot}
	Time evolution of the QBER for the 1533/1577 signal/idler set.  
	The horizontal errorbars denote the acquisition time for each run (usually 500 seconds), 
	while the vertical bars represent the QBER uncertainty (1 standard deviation) due to statistical fluctuations.  	
	}
\end{figure}

The temporal stability of the system is also tested.  
Using the 1533/1577 frequency-conjugate set, the QKD system is allowed to run many times over the course of 
two hours (7200 seconds), with no adjustments made to the system over this time.  
The QBER ($\pm \sigma$) in each run is measured and plotted in Fig. \ref{fig:QBErplot}; the horizontal error bars represent the acquisition time $T_{\textrm{acq}}$ ($\approx 500$ s) of the run.  We observe that over the course of two hours, the QBER does not change dramatically, implying that the polarization does not drift 
significantly in a tabletop laboratory setting even over a 40 km spool of fiber with no active polarization stabilization.  
A total sifted key of 59.4 kbits and an estimated secure key size (Eqn. \ref{eq:Rsec}) of 31.2 kbits is generated 
after this two-hour time period.

We observe (Table \ref{tab:QKDresults}) that secure key rates of up to 20.5  bits/s can be generated.  
This key rate, however, can be significantly improved upon if:
\begin{itemize}
\item Alice and Bob each have a 
second detector, with one detector measuring in the $H/V$ basis and the other detector the $D/A$ basis ($\times 4$), 
\item fiber splice losses due to poled-fiber and other fiber-coupled devices were reduced ($\times 4$),
\item the detector efficiencies were to increase from the current $\sim 20\%$ to 
$60\%$ ($\times 9$) by utilizing practical avalanche-photodiode (APD)-based SPDs \cite{SPD50percent,restelli2013single}, and
\item the pump laser repetition rate were increased from 81.6 MHz to 2 GHz ($\times 25$), 
generating photon pairs that can be detected using low-deadtime APD-based self-differencing \cite{SPD50percent} or harmonically-gated  \cite{restelli2013single} SPDs.  
\end{itemize}
These improvements would result in a 3600-fold enhancement of the secure-key rate to 75 kbits/s for each signal/idler channel, or an aggregate
1.5 Mbits/s when the entire spectrum of the poled fiber is utilized.  
In addition, the strong spectral correlations of our entangled source (Fig. \ref{fig:PPSF2014_JSA}) can be exploited for high-dimensional QKD \cite{Nunn:13}, with such a scheme enabling the generation of many more than 1 secure bit per coincidence detection \cite{zhong2015photon}.  
\newline

In summary, we have described a scheme for the distribution of broadband entangled photon pairs 
to multiple users using standard DWDM wavelength channels.  
A proof-of-principle QKD experiment for this scheme was performed, yielding estimated secure key rates of up to 20 bits/s per channel.  
This brings us a step closer toward an entanglement-based, reconfigurable multi-user quantum network that is compatible (without additional resources) to current optical network infrastructures.

\section*{Funding}
We acknowledge the Natural Sciences and Engineering Research Council (Canada)
 and the EU Project CHARMING (Contract No. FP7-288786)  for 
funding the work presented in this paper.



%

\end{document}